\documentclass[conference]{IEEEtran}
\usepackage{cite}

\ifCLASSINFOpdf
\else
\fi
\usepackage[cmex10]{amsmath}
\usepackage{amssymb}

\usepackage{tikz}
\usepackage{pgfplots}
\pgfplotsset{compat=1.3}
\usetikzlibrary{arrows,snakes,shapes,positioning,arrows,decorations.markings,calc}

\DeclareMathAlphabet{\mathbit}{OML}{cmr}{bx}{it}

\DeclareMathOperator{\Q}{Q}
\DeclareMathOperator{\E}{E}

\DeclareMathOperator{\Diag}{diag}

\DeclareMathOperator{\fieldR}{\mathbb{R}}

\newcommand{\ve}[1]{\boldsymbol{#1}}

\newcommand{\ex}[1]{\E \left[#1\right]}

\newcommand{\diag}[1]{\Diag \left(#1\right)}

\newcommand{\qfunc}[1]{\Q \left(#1\right)}

\begin{document}
%
\title{Overdemodulation for High-Performance\\Receivers with Low-Resolution ADC}
%
%
%
\author{
\IEEEauthorblockN{Manuel Stein, Sebastian Theiler and Josef A. Nossek}
Technische Universit\"at M\"unchen, Munich, Germany 
\\ Email: \{manuel.stein, sebastian.theiler, josef.a.nossek\}@tum.de
}

\maketitle



%
\IEEEpeerreviewmaketitle

%
%
%
%
\begin{abstract}
The design of the analog demodulator for receivers with low-resolution analog-to-digital converters (ADC) is investigated. For infinite ADC resolution, demodulation to baseband with $M=2$ orthogonal sinusoidal functions (quadrature demodulation) is an optimum design choice. For receive systems which are restricted to ADC with low amplitude resolution we show here that this classical demodulation approach is suboptimal. To this end we analyze the theoretical channel parameter estimation performance based on a simple pessimistic characterization of the Fisher information measure when forming $M>2$ analog demodulation channels prior to an ADC with $1$-bit amplitude resolution. In order to emphasize that this inside is also true for communication problems, we provide an additional discussion on the behavior of the Shannon information measure under overdemodulation and $1$-bit quantization.
\end{abstract}
\section{Introduction}
For the design of future portable wireless receivers,  the design of the ADC has been identified as one of the bottlenecks when heading at a receiver architecture achieving an optimum trade-off between low cost, moderate energy consumption and high performance \cite{Walden99}. As the complexity and the power dissipation of the ADC scales exponentially $\mathcal{O}(2^b)$ with the bits $b$ used for amplitude resolution, receiver designs with low ADC resolution, e.g. $1$-bit amplitude resolution, have recently gained growing attention \cite{Host00} \cite{Dabeer06} \cite{koch1}. While such an ADC design is highly attractive with respect to complexity, it degrades the achievable performance of the receiver. Interestingly, for applications with low signal-to-noise ratio (SNR) the relative performance gap between a symmetric hard-limiting $1$-bit system and an ideal receiver with infinite ADC resolution is moderate with $2/\pi$ ($-1.96$ dB) \cite{Vleck66}. In contrast, for the medium to high SNR regime the loss is quiet pronounced. Recently, the potential of different techniques for the reduction of the quantization-loss is discussed in various works. The benefit of oversampling the analog receive signal for communication over a noisy channel is discussed in \cite{koch1}  \cite{Krone12}. In \cite{koch2} the authors analyze the adjustment of the quantization threshold. The work \cite{Mez12} observes that noise correlation can increase the capacity of multiple-input multiple-output (MIMO) communication channels with coarse receive quantization, while \cite{SteinICASSP13} shows that adjusting the analog pre-filter prior to a $1$-bit ADC partially recovers the quantization-loss for channel estimation problems. In this context here the design of the demodulation device prior to a low-resolution ADC is discussed. In order to demodulate the carrier signal of each sensor to baseband, classical receivers use a demodulator with in-phase and quadrature channel. Within each channel the receive signal is multiplied by a sinusoidal function oscillating at carrier frequency, where the two functions are kept orthogonal through a phase offset of $\frac{\pi}{2}$ \cite[p. 582ff.]{Oppenheim96}. While for receivers with infinite ADC resolution this method induces no information-loss during the transition from the analog to the digital domain, here we show that using $M>2$ analog demodulation channels prior to an ADC of low resolution allows to significantly reduce the loss due to $1$-bit quantization.
\section{System Model}
For the discussion we assume a transmitter sending
\begin{align}
x'(t)=x_1'(t)\sqrt{2} \cos{(\omega_c t)}-x_2'(t)\sqrt{2} \sin{(\omega_c t)},
\end{align}
where $\omega_c\in\fieldR$ is the carrier frequency and $x_{1/2}'(t)\in\fieldR$ are two independent input signals. The analog receiver observes
\begin{align}
y'(t)&=\gamma x_1'(t-\tau)\sqrt{2} \cos{(\omega_c t - \phi)}-\notag\\ 
&\quad\quad\quad-\gamma x_2'(t-\tau)\sqrt{2} \sin{(\omega_c t - \phi)}+\eta'(t),
\end{align}
where $\gamma\in\fieldR, \gamma\geq0$ is the attenuation and $\tau\in\fieldR$ a time-shift due to signal propagation. $\phi\in\fieldR$ characterizes the channel phase offset and $\eta'(t)\in\fieldR$ is white additive sensor noise. For the demodulation to baseband the receiver forms $m=1,\ldots,M$ channel outputs by performing the multiplications
\begin{align}
y_m'(t)&=y'(t) \cdot \sqrt{2} \cos{(\omega_c t + \varphi_m)} \notag\\
&=\gamma x_1'(t-\tau) \big(\cos{(2\omega_c t - \phi + \varphi_m)} + \cos{(\phi + \varphi_m)} \big) -  \notag\\
&-\gamma x_2'(t-\tau) \big(\sin{(2\omega_c t - \phi + \varphi_m)} - \sin{(\phi + \varphi_m)} \big) +  \notag\\
&+\eta'(t)\sqrt{2} \cos{(\omega_c t + \varphi_m)},
\end{align}
with demodulation offsets $\varphi_m$. Behind a low-pass filter $h(t;B)$ of bandwidth $B$, the $m$-th output channel is
\begin{align}
y_m(t)&=\gamma x_1(t-\tau) \big( \cos{(\phi)}\cos{(\varphi_m)} - \sin{(\phi)}\sin{(\varphi_m)} \big)+\notag\\
&+\gamma x_2(t-\tau) \big( \sin{(\phi)}\cos{(\varphi_m)} + \cos{(\phi)}\sin{(\varphi_m)} \big)+\notag\\
&+\cos{(\varphi_m)}\eta_1(t)+\sin{(\varphi_m)}\eta_2(t),
\end{align}
where 
\begin{align}
\eta_1(t)&=\sqrt{2}\cos{(\omega_c t )} \big(h(t;B)*\eta'(t)\big)\notag\\
\eta_2(t)&= - \sqrt{2}\sin{(\omega_c t )}\big(h(t;B)*\eta'(t)\big)\label{noise:demodulation}
\end{align}
are two independent random processes with power spectral density $\Phi(\omega)=1$. Note that we use the convention $z(t)=h(t;B)*z'(t)$, where $*$ is the convolution operator. Defining the demodulation offset vector
\begin{align}
\ve{\varphi}=\begin{bmatrix} \varphi_1 &\varphi_2 &\ldots &\varphi_M \end{bmatrix}^{\rm T},
\end{align}
the signals of the $M$ demodulation channels can be written
\begin{align}
\ve{y}(t)=\ve{A}(\ve{\varphi})\big(\gamma \ve{B}(\phi)\ve{x}(t-\tau) + \ve{\eta}(t)\big), \label{model:matrix:vector}
\end{align}
with the analog signals
\begin{align}
\ve{y}(t)&=\begin{bmatrix} y_1(t) &y_2(t) &\ldots &y_M(t) \end{bmatrix}^{\rm T}\notag\\
\ve{x}(t-\tau)&=\begin{bmatrix} x_1(t-\tau) &x_2(t-\tau)\end{bmatrix}^{\rm T}\notag\\
\ve{\eta}(t)&=\begin{bmatrix} \eta_1(t) &\eta_2(t)\end{bmatrix}^{\rm T}
\end{align}
and the matrices
\begin{align}
\ve{A}(\ve{\varphi})&=
\begin{bmatrix} 
\cos{(\varphi_1)} & \sin{(\varphi_1)}\\  
\cos{(\varphi_2)} & \sin{(\varphi_2)}\\  
\vdots & \vdots\\  
\cos{(\varphi_M)} & \sin{(\varphi_M)}\\  
\end{bmatrix}\notag\\
\ve{B}(\phi)&=
\begin{bmatrix} 
\cos{(\phi)} & \sin{(\phi)}\\  
-\sin{(\phi)} & \cos{(\phi)}\\  
\end{bmatrix}.
\end{align}
Sampling each of the $M$ output channels at a rate of $f_s=\frac{1}{T_s}=2B$ for the duration of $T=NT_s$ and defining the parameter vector $\ve{\theta}=\begin{bmatrix} \phi &\tau \end{bmatrix}^{\rm T}$, the digital receive signal is comprised by $N$ temporally white snapshots $\ve{y}_n\in\fieldR^M$ with
\begin{align}
\ve{y}_n&=\gamma\ve{A}(\ve{\varphi})\ve{B}(\phi)\ve{x}_n(\tau) + \ve{A}(\ve{\varphi})\ve{\eta}_n\notag\\
&=\gamma \ve{s}_n(\ve{\theta})+\ve{\zeta}_n.
\end{align}
The individual digital samples are
\begin{align}
\ve{y}_n&=\begin{bmatrix} y_1\Big(\frac{(n-1)}{f_s}\Big) &y_2\Big(\frac{(n-1)}{f_s}\Big) &\ldots &y_M\Big(\frac{(n-1)}{f_s}\Big) \end{bmatrix}^{\rm T}\notag\\
\ve{x}_n(\tau)&=\begin{bmatrix} x_1\Big(\frac{(n-1)}{f_s}-\tau\Big) &x_2\Big(\frac{(n-1)}{f_s}-\tau\Big)\end{bmatrix}^{\rm T}\notag\\
\ve{\eta}_n&=\begin{bmatrix} \eta_1\Big(\frac{(n-1)}{f_s}\Big) &\eta_2\Big(\frac{(n-1)}{f_s}\Big)\end{bmatrix}^{\rm T}.
\end{align}
The sampled noise $\ve{\eta}_n$ is a zero-mean Gaussian variable with $\ex{\ve{\eta}_n\ve{\eta}_n^{\rm T}}=\ve{I}_2$ and the noise covariance matrix for each snapshot is given by
\begin{align}
\ve{C}&=\ex{\ve{\zeta}_n\ve{\zeta}^{\rm T}_n}
=\ve{A}(\ve{\varphi})\ve{A}^{\rm T}(\ve{\varphi}).
\end{align}
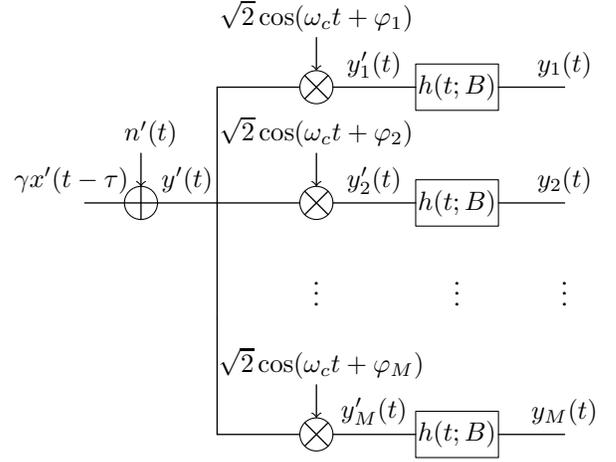
\begin{figure}
\begin{center}
\begin{tikzpicture}[line cap=round,line join=round,yscale=0.44, xscale=0.44]
   
  	\draw[color=black,line width=0.5] (-2,5.5) -- (2,5.5);
	\draw (-2.4,5.5)[above] node {$\gamma x'(t-\tau)$};
	
  	\draw (-0.3,5.5) circle (0.5cm);
  	\draw[color=black,line width=0.5] (-0.3,5) -- (-0.3,6);
	\draw[->,color=black, line width=0.5] (-0.3,7) -- (-0.3,6);
  	\draw (0,6.9)[above] node {$n'(t)$};
	
	\draw (1.1,5.5)[above] node {$y'(t)$};
  
  	\draw[color=black,line width=0.5] (2,5.5) -- (2,9);
  	\draw[color=black,line width=0.5] (2,5.5) -- (2,2);
	\draw[color=black,line width=0.5] (2,2) -- (2,-1.5);
  
  	\draw[color=black,line width=0.5] (2,9) -- (4.5,9);
	\draw[color=black,line width=0.5] (2,5.5) -- (4.5,5.5);
  	\draw[color=black,line width=0.5] (2,-1.5) -- (4.5,-1.5);

 	\draw (5,9) circle (0.5cm);
	\draw (5,5.5) circle (0.5cm);
  	\draw (5,-1.5) circle (0.5cm);

  	\draw[color=black,line width=0.5] (5-0.35,9+0.35) -- (5+0.35,9-0.35);
  	\draw[color=black,line width=0.5] (5+0.35,9+0.35) -- (5-0.35,9-0.35);
	
	 \draw[color=black,line width=0.5] (5-0.35,5.5+0.35) -- (5+0.35,5.5-0.35);
  	\draw[color=black,line width=0.5] (5+0.35,5.5+0.35) -- (5-0.35,5.5-0.35);
  
  	\draw[color=black,line width=0.5] (5-0.35,-1.5+0.35) -- (5+0.35,-1.5-0.35);
  	\draw[color=black,line width=0.5] (5+0.35,-1.5+0.35) -- (5-0.35,-1.5-0.35);
  
  	\draw[->,color=black, line width=0.5] (5,10.5) -- (5,9.5);
  	\draw (5,10.4) node [above] {$\sqrt{2}\cos(\omega_c t+\varphi_1)$};
	
	\draw[->,color=black, line width=0.5] (5,7.0) -- (5,6.0);
  	\draw (5,6.9) node [above] {$\sqrt{2}\cos(\omega_c t+\varphi_2)$};

	\draw[->,color=black, line width=0.5] (5,0) -- (5,-1);
	\draw (5.15,-0.1) node [above] {$\sqrt{2}\cos(\omega_c t+\varphi_M)$};
	
  	\draw[color=black,line width=0.5] (5.5,9) -- (8,9);
	\draw[color=black,line width=0.5] (5.5,5.5) -- (8,5.5);
  	\draw[color=black,line width=0.5] (5.5,-1.5) -- (8,-1.5);

	\draw (6.75,9)[above] node {$y'_1(t)$};
	\draw (6.75,5.5)[above] node {$y'_2(t)$};
	\draw (6.75,-1.5)[above] node {$y'_M(t)$};
	
	\draw (8,8.3) rectangle (10.5,9.7);
	\draw (8,4.8) rectangle (10.5,6.2);
  	\draw (8,-2.2) rectangle (10.5,-0.8);

	\draw (9.25,9) node {$h(t;B)$};
	\draw (9.25,5.5) node {$h(t;B)$};
	\draw (9.25,-1.5) node {$h(t;B)$};
  
  	\draw[color=black,line width=0.5] (10.5,9) -- (12.5,9);
	\draw[color=black,line width=0.5] (10.5,5.5) -- (12.5,5.5);
  	\draw[color=black,line width=0.5] (10.5,-1.5) -- (12.5,-1.5);
	
	\draw (12.5,9)[above] node {$y_1(t)$};
	\draw (12.5,5.5)[above] node {$y_2(t)$};
	\draw (12.5,-1.5)[above] node {$y_M(t)$};
	
	\draw (5,3) node {$\vdots$};
	\draw (9.25,3) node {$\vdots$};
	\draw (12.5,3) node {$\vdots$};

\end{tikzpicture}
\end{center}
\vspace{-0.2cm}
\caption{Analog radio front-end design with overdemodulation}
\label{fig:classical}
\end{figure}
In the following we assume, that the ADC for each of the $M$ output channels is restricted to a symmetric hard-limiter, such that the final digital receive data $\ve{r}_n\in\{-1,1\}^M$ is given by
\begin{align}
\ve{r}_n=\operatorname{sign}\big(\ve{y}_n\big),
\end{align}
where $\operatorname{sign}(\cdot)$ is the element-wise signum-function.
\section{Performance Analysis - Estimation}
In order to discuss the benefits of using $M>2$ demodulation outputs, a channel estimation problem is considered. The receiver infers the deterministic but unknown parameters $\ve{\theta}$, by using the maximum-likelihood estimator (MLE)
\begin{align}
\ve{\hat{\theta}}(\ve{r})&=\arg\max_{\ve{\theta}\in\Theta} \ln p(\ve{r};\ve{\theta}),
\end{align}
where the receive signal with $N$ snapshots has the form
\begin{align}
\ve{r}=\begin{bmatrix} \ve{r}_1^{\rm T} &\ve{r}_2^{\rm T} &\ldots &\ve{r}_N^{\rm T} \end{bmatrix}^{\rm T}.
\end{align}
For sufficiently large $N$, the MLE is unbiased and efficient, such that its asymptotic MSE matrix $\ve{\bar{R}}_{\ve{\hat{\theta}}}$ can be characterized analytically through the Cram\'er-Rao lower bound \cite{Kay93}, given by the inverse of the Fisher information matrix (FIM)
\begin{align}
\ve{\bar{R}}_{\ve{\hat{\theta}}}&=\lim_{N\to\infty} \ex{ (\ve{\hat{\theta}}(\ve{r})-\ve{\theta})(\ve{\hat{\theta}}(\ve{r})-\ve{\theta})^{\rm T} }\notag\\
&=\ve{F}^{-1}(\ve{\theta}).
\end{align}
The FIM is defined by
\begin{align}
\ve{F}(\ve{\theta})=\int_{\mathcal{R}}  p(\ve{r};\ve{\theta}) \Bigg(\frac{\partial \ln{p(\ve{r};\ve{\theta})}}{\partial\ve{\theta}} \Bigg)^{\rm T} \frac{\partial \ln{p(\ve{r};\ve{\theta})}}{\partial\ve{\theta}}  \mathrm d\ve{r},
\end{align}
where $\mathcal{R}$ is the support of the receive vector $\ve{r}$. For temporally white samples $\ve{r}_n$, the FIM $\ve{F}(\ve{\theta})$ exhibits an additive property
\begin{align}
\ve{F}(\ve{\theta})&=\sum_{n=1}^{N} \ve{F}_n(\ve{\theta})\notag\\
\ve{F}_n(\ve{\theta})&=\int_{\mathcal{R}_n} p(\ve{r}_n;\ve{\theta}) \Bigg(\frac{\partial \ln{p(\ve{r}_n;\ve{\theta})}}{\partial\ve{\theta}}  \Bigg)^2 \mathrm d \ve{r}_n.\label{measure:fisher:sample}
\end{align}
As (\ref{measure:fisher:sample}) requires in general the calculation of an $M$-fold integral, the analytic description of $\ve{F}_n(\ve{\theta})$ is difficult, especially if $M$ is large. To circumvent this problem we use an approximation $\ve{\tilde{F}}_n(\ve{\theta})$ of the FIM which exhibits the property
\begin{align}
\ve{F}_n(\ve{\theta}) \succeq \ve{\tilde{F}}_n(\ve{\theta}).
\end{align}
This guarantees that $\ve{\tilde{F}}_n(\ve{\theta})$ is a pessimistic characterization of the performance measure $\ve{F}_n(\ve{\theta})$. With the moments
\begin{align}
\ve{\mu}_n(\ve{\theta})&=\int_{\mathcal{R}_n} \ve{r}_n p(\ve{r}_n;\ve{\theta}) \mathrm d\ve{r}_n\notag\\
\ve{R}_n(\ve{\theta})&=\int_{\mathcal{R}_n} \big(\ve{r}_n-\ve{\mu}_n(\ve{\theta})\big)\big(\ve{r}_n-\ve{\mu}_n(\ve{\theta})\big)^{\rm T}  p(\ve{r}_n;\ve{\theta}) \mathrm d \ve{r}_n,
\end{align}
such a pessimistic version of the FIM is given by \cite{SteinARXIV13}
\begin{align}
\ve{\tilde{F}}_n(\ve{\theta}) = \bigg(\frac{\partial \ve{\mu}_n(\ve{\theta}) }{\partial\ve{\theta}} \bigg)^{\rm T} \ve{R}_n^{-1} (\ve{\theta})\bigg(\frac{\partial \ve{\mu}_n(\ve{\theta}) }{\partial\ve{\theta}} \bigg).
\end{align}
The first moment can be calculated element-wise by
\begin{align}
[\ve{\mu}_{n}(\ve{\theta})]_m&=p\big([\ve{r}_{n}]_m =1; \ve{\theta}\big)-p\big([\ve{r}_{n}]_m = -1; \ve{\theta}\big)\notag\\
&=1-2\qfunc{ \frac{ \gamma [\ve{s}_{n}(\ve{\theta})]_m}{\sqrt{[\ve{C}]_{mm}}}},
\end{align}
where $\qfunc{\cdot}$ is the Q-function. Further, the second moment
\begin{align}
[\ve{R}_{n}(\ve{\theta})]_{mm}&=1-[\ve{\mu}_{n}(\ve{\theta})]_m^2,
\end{align}
with off-diagonal entries
\begin{align}
[\ve{R}_{n}(\ve{\theta})]_{mk}
&=4\Phi_{mk}(\ve{\theta}) -\big(1-[\ve{\mu}_{n}(\ve{\theta})]_m\big) \big(1-[\ve{\mu}_{n}(\ve{\theta})]_k\big),
\end{align}
where $\Phi_{mk}(\ve{\theta}) $ is the cumulative density function (CDF) of the bivariate Gaussian distribution
\begin{align}
p\big([\ve{\zeta}_{n}]_m,[\ve{\zeta}_{n}]_k\big)
&=\mathcal{N}\Bigg(\begin{bmatrix} 0\\ 0 \end{bmatrix},\begin{bmatrix} [\ve{C}]_{mm} &[\ve{C}]_{mk}\\ [\ve{C}]_{km} &[\ve{C}]_{kk} \end{bmatrix} \Bigg)
\end{align}
with upper integration boarder $\begin{bmatrix} -\gamma [\ve{s}_{n}(\ve{\theta})]_m -\gamma [\ve{s}_{n}(\ve{\theta})]_k\end{bmatrix}^{\rm T}$. The derivative of the first moment is found element-wise by
\begin{align}
\Bigg[ \frac{\partial \ve{\mu}_n(\ve{\theta}) }{\partial\ve{\theta}} \Bigg]_{mk}= \frac{2 \gamma \mathrm e^{- \frac{ \gamma^2 [\ve{s}_{n}(\ve{\theta})]_m^2}{{2 [\ve{C}]_{mm}}} } }{ \sqrt{2 \pi [\ve{C}]_{mm} } } \Bigg[ \frac{\partial \ve{s}_n(\ve{\theta}) }{\partial\ve{\theta}} \Bigg]_{mk},
\end{align}
with
\begin{align}
\frac{\partial \ve{s}_n(\ve{\theta}) }{\partial\ve{\theta}} &= \begin{bmatrix} \frac{\partial \ve{s}_n(\ve{\theta}) }{\partial\phi} &\frac{\partial \ve{s}_n(\ve{\theta}) }{\partial\tau} \end{bmatrix}\notag\\
&= \begin{bmatrix} \ve{A}(\ve{\varphi}) \frac{\partial \ve{B}(\phi)}{\partial \phi} \ve{x}_n(\tau)  &\ve{A}(\ve{\varphi})\ve{B}(\phi) \frac{\partial \ve{x}_n(\tau)}{\partial \tau}\end{bmatrix},
\end{align}
where
\begin{align}
\frac{\partial \ve{B}(\phi)}{\partial \phi}&=
\begin{bmatrix} 
-\sin{(\phi)} & \cos{(\phi)}\\  
-\cos{(\phi)} & -\sin{(\phi)}\\  
\end{bmatrix}\notag\\
\frac{\partial \ve{x}_n(\tau)}{\partial \tau}&=-\begin{bmatrix} \frac{\mathrm d x_1(t)}{\mathrm d t}&\frac{\mathrm d x_2(t)}{\mathrm d t} \end{bmatrix}^{\rm T}\Big|_{t=\big(\frac{(n-1)}{f_s}-\tau\big)} .
\end{align}
\subsection{Results - Channel Estimation}
For visualization of the possible performance gain we use an example where the transmitter sends pilot signals
\begin{align}
x_{1/2}(t)&= \sum_{k=-\infty}^{\infty} [\ve{b}_{1/2}]_{\operatorname{mod}{(k,K)}} g(t-kT_{b}).
\end{align}
$\ve{b}_{1/2} \in \{-1, 1\}^{K}$ are binary vectors with $K=1023$ symbols, each of duration $T_{b}=977.52$ ns, $g(t)$ is a rectangular transmit pulse and $\operatorname{mod}{(\cdot)}$ is the modulo operator. The receiver band-limits the signal to $B=1.023$ MHz and samples at a rate of $f_s=2 B$ in order to attain temporally white snapshots. After one signal period $T=1$ ms, the receiver has available $N=2046$ samples for the estimation task. The unknown channel parameters are assumed to be $\ve{\theta}=\begin{bmatrix} \frac{\pi}{8} &0\end{bmatrix}^{\rm T}$. The $M$ demodulation offsets are equally spaced $[\ve{\varphi}]_m=\frac{\pi}{M}(m-1)$ and the performance is measured in relation to an ideal reference system with infinite ADC resolution and $M=2$
\begin{align}
\chi_{\phi/\tau}(\ve{\theta})&= \frac{[\ve{\tilde{F}}^{-1}(\ve{\theta})]_{11/22}}{[\ve{F}_{\infty}^{-1}(\ve{\theta})]_{11/22}},
\end{align}
where the FIM of the reference system is
\begin{align}
\ve{F}_{\infty}(\ve{\theta})=  \gamma^2 \sum_{n=1}^{N} \bigg(\frac{\partial \ve{s}_n(\ve{\theta}) }{\partial\ve{\theta}} \bigg)^{\rm T}\bigg(\frac{\partial \ve{s}_n(\ve{\theta}) }{\partial\ve{\theta}} \bigg).
\end{align}
Note, that for $M=2$ the noise in both demodulation channels is independent. Under this condition it holds that the approximated FIM with hard-limiting is exact \cite{SteinARXIV13}, i.e. $\ve{\tilde{F}}(\ve{\theta})=\ve{{F}}(\ve{\theta})$. Therefore, $\chi_{\phi/\tau}(\ve{\theta})\big|_{M=2}$ characterizes the $1$-bit performance loss with classical I/Q demodulation precisely. For the case $M>2$ the ratio $\chi_{\phi/\tau}(\ve{\theta})$ provides a pessimistic approximation, i.e. the quantization-loss might even be smaller. Fig. \ref{fig1} and \ref{fig2} show the estimation performance $\chi_{\phi}(\ve{\theta})$ and $\chi_{\tau}(\ve{\theta})$ for different choices of $M$ versus signal-to-noise ratio (SNR). For both parameters overdemodulation with $M=16$ allows to diminish the quantization-loss at $\operatorname{SNR}=-15.0$ dB from $\chi_{\phi/\tau}(\ve{\theta})=-1.99$ dB to $\chi_{\phi/\tau}(\ve{\theta})=-1.07$ dB. For high SNR scenarios, the gain is especially pronounced for the estimation of the phase parameter $\phi$. At $\operatorname{SNR}=+10.0$ dB the loss can be reduced from $\chi_{\phi}(\ve{\theta})=-7.92$ dB to $\chi_{\phi}(\ve{\theta})=-0.51$ dB. For the time-delay parameter $\tau$, the $1$-bit loss at $\operatorname{SNR}=+10.0$ dB reduces from $\chi_{\tau}(\ve{\theta})=-6.45$ dB to $\chi_{\tau}(\ve{\theta})=-3.18$ dB.
%
\pgfplotsset{legend style={rounded corners=2pt,nodes=right}}
\begin{figure}[!htbp]
\begin{tikzpicture}

  	\begin{axis}[ylabel=$\chi_{\phi}(\ve{\theta})\text{ in dB}$,
  			xlabel=$\text{$\operatorname{SNR}$ in dB}$,
			grid,
			ymin=-6.0,
			ymax=0.0,
			xmin=-15,
			xmax=10,
			legend pos=south west]
			
			\addplot[black, style=densely dashed] table[x index=0, y index=5]{Fisher_1bit_M16.txt};
			\addlegendentry{$M=16$}
			
			\addplot[green!60!black, style=solid] table[x index=0, y index=5]{Fisher_1bit_M8.txt};
			\addlegendentry{$M=8$}
			
			\addplot[red, style=solid] table[x index=0, y index=5]{Fisher_1bit_M5.txt};
			\addlegendentry{$M=5$}
			
			\addplot[blue, style=solid] table[x index=0, y index=5]{Fisher_1bit_M3.txt};
			\addlegendentry{$M=3$}
			
			\addplot[black, style=solid] table[x index=0, y index=4]{Fisher_1bit_M3.txt};
			\addlegendentry{$M=2$}

	\end{axis}
	
\end{tikzpicture}
\vspace{-0.25cm}
\caption{$\chi_{\phi}(\ve{\theta})$ vs. signal-to-noise ratio $\operatorname{SNR}$}
\label{fig1}
\end{figure}
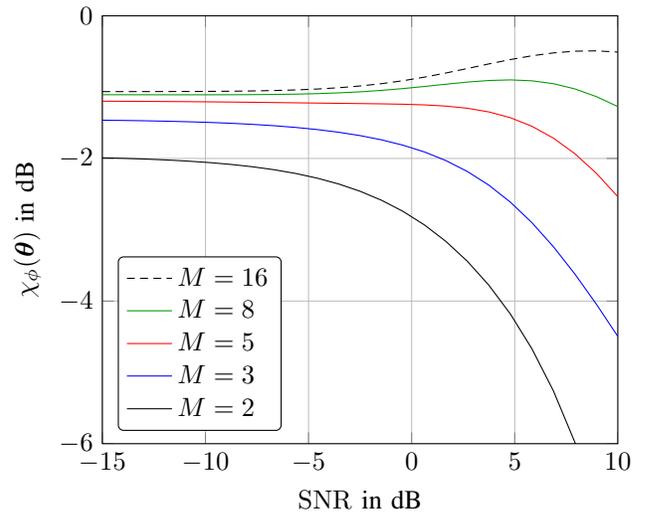
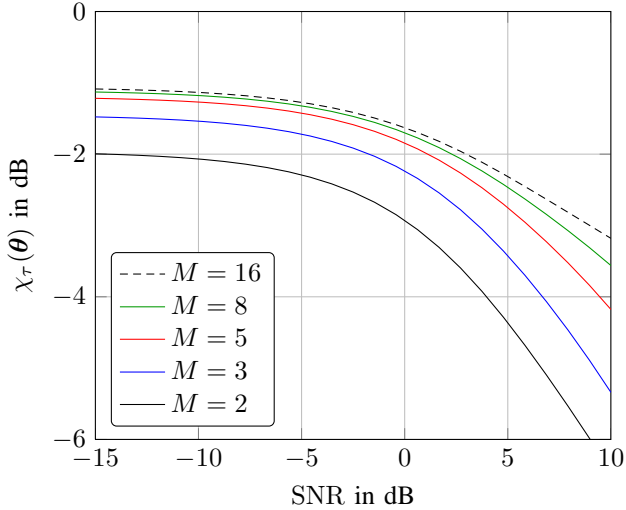
\begin{figure}[!htbp]
\begin{tikzpicture}

  	\begin{axis}[ylabel=$\chi_{\tau}(\ve{\theta})\text{ in dB}$,
  			xlabel=$\text{$\operatorname{SNR}$ in dB}$,
			grid,
			ymin=-6.0,
			ymax=0.0,
			xmin=-15,
			xmax=10,
			legend pos=south west]
			
			\addplot[black, style=densely dashed] table[x index=0, y index=2]{Fisher_1bit_M16.txt};
			\addlegendentry{$M=16$}
			
			\addplot[green!60!black, style=solid] table[x index=0, y index=2]{Fisher_1bit_M8.txt};
			\addlegendentry{$M=8$}
			
			\addplot[red, style=solid] table[x index=0, y index=2]{Fisher_1bit_M5.txt};
			\addlegendentry{$M=5$}
			
			\addplot[blue, style=solid] table[x index=0, y index=2]{Fisher_1bit_M3.txt};
			\addlegendentry{$M=3$}
			
			\addplot[black, style=solid] table[x index=0, y index=1]{Fisher_1bit_M3.txt};
			\addlegendentry{$M=2$}

	\end{axis}
	
\end{tikzpicture}
\vspace{-0.25cm}
\caption{$\chi_{\tau}(\ve{\theta})$ vs. signal-to-noise ratio $\operatorname{SNR}$}
\label{fig2}
\end{figure}
\section{Performance Analysis - Communication}
In the context of communication theory, our setup can be interpreted as a real-valued multiple-input and multiple-output (MIMO) channel with two inputs and $M$ channel outputs
\begin{align}
\ve{y}&=\ve{A}(\ve{\varphi})\ve{B}(\phi)\ve{x}+ \ve{A}(\ve{\varphi})\ve{\eta}=\ve{H}\ve{x}+\ve{\zeta},
\end{align}
followed by an element-wise hard-limiter $\ve{r}=\operatorname{sign}(\ve{y})$. For such a channel it was shown in \cite{Mez12} that the Shannon information measure $I(\ve{x};\ve{r})$, related to the maximum achievable transmission rate, can be approximated from below by
\begin{align}
I(\ve{x};\ve{r})&\geq \frac{1}{2} \log_2 \det{\Big(\ve{1}_M+\ve{R}_{\zeta'\zeta'}^{-1}\ve{H}'\ve{R}_{xx}\ve{H}'^{\rm T}\Big)}\notag\\
&=\tilde{I}(\ve{x};\ve{r}),\label{bound:capacity}
\end{align}
where $\ve{R}_{xx}$ is the second moment of the channel input $\ve{x}$ and
\begin{align}
\ve{R}_{\zeta'\zeta'}&=\frac{2}{\pi} \Big( \arcsin{ \big(\diag{\ve{R}_{yy}}^{-\frac{1}{2}}\ve{R}_{yy}\diag{\ve{R}_{yy}}^{-\frac{1}{2}} \big)} \Big)\notag\\
&-\frac{2}{\pi}  \diag{\ve{R}_{yy}}^{-\frac{1}{2}}\ve{R}_{yy}\diag{\ve{R}_{yy}}^{-\frac{1}{2}}\notag\\
&+\frac{2}{\pi} \diag{\ve{R}_{yy}}^{-\frac{1}{2}}\ve{R}_{\zeta\zeta}\diag{\ve{R}_{yy}}^{-\frac{1}{2}}\notag\\
\ve{H}'&=\sqrt{\frac{2}{\pi}} \diag{\ve{R}_{yy}}^{-\frac{1}{2}}\ve{H}.
\end{align}
Note that with $M=2$ and $1$-bit quantization the capacity of the considered transmission line is \cite{Dabeer06}
\begin{align}
C&=\max_{p(\ve{x})} I(\ve{x};\ve{r})=2\Big(1-\beta\Big(\qfunc{\sqrt{\operatorname{SNR}}}\Big)\Big)
\end{align}
with $\beta(z)=-z\log_2(z)-(1-z)\log_2(1-z)$.
\subsection{Results - Noisy Channel Communication}
\begin{figure}[!htbp]
\begin{tikzpicture}

  	\begin{axis}[ylabel=${\tilde{I}(\ve{x};\ve{r})}/{C}$,
  			xlabel=$M$,
			grid,
			ymin=0.95,
			ymax=1.25,
			xmin=2,
			xmax=20,
			legend pos=south east]
			
			\addplot[red, style=solid] table[x index=0, y index=1]{Capacity_NumM_15dB.txt};
			\addlegendentry{$\operatorname{SNR}=-15.0\text{ dB}$}

	\end{axis}
	
\end{tikzpicture}
\vspace{-0.25cm}
\caption{Transmission-rate $\chi$ vs. demodulation channels $M$}
\label{fig3}
\end{figure}
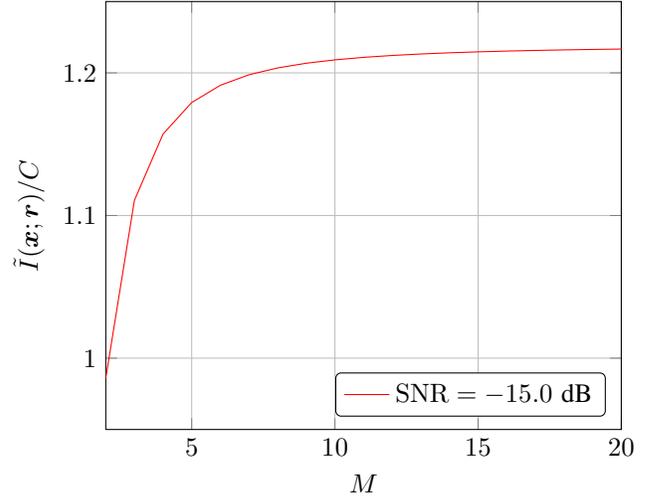

For simulations we assume independent channel inputs with zero-mean and covariance $\ve{R}_{xx}=\operatorname{SNR}\cdot\ve{I}_2$. Fig. \ref{fig3} shows the achievable gain in transmission rate with $1$-bit ADC at the receiver and different numbers of demodulation channels $M$. It is observed that classical demodulation (quadrature demodulation) is suboptimal as with overdemodulation ($M=20$) it is possible to increase the transmission rate by $22\%$ in a low SNR scenario with $\operatorname{SNR}=-15.0$ dB.
\section{Conclusion}
A receiver design which uses $M>2$ analog demodulation channels to map the carrier signal to baseband has been analyzed. While for receivers with high ADC resolution this approach leads to redundant receive data and therefore has no advantage, here it was shown by an estimation and communication theoretic investigation, that for receivers which are restricted to ultra low ADC resolution significant performance improvements can be achieved if more than two demodulation channels are used. Key to this gain is to create redundancy before passing the signal through a highly non-linear device.
\ifCLASSOPTIONcaptionsoff
  \newpage
\fi
\end{document}